\documentclass[aps,prd,preprint,tightenlines,showpacs,nofootinbib,showkeys]{revtex4}

\usepackage{epsfig}
\usepackage{graphicx}

\begin{document}
\title{Correspondence between Einstein-Yang-Mills-Lorentz systems and dynamical torsion models}

\author{Jose A. R. Cembranos$^{a,b}$\footnote{E-mail: cembra@fis.ucm.es},
and Jorge Gigante Valcarcel$^{a}$\footnote{E-mail: jorgegigante@ucm.es}
}

\affiliation{$^{a}$Departamento de  F\'{\i}sica Te\'orica I, Universidad Complutense de Madrid, E-28040 Madrid, Spain\\
$^{b}$Departamento de  F\'{\i}sica, Universidade de Lisboa,
P-1749-016 Lisbon, Portugal
}




%
%
%

\begin{abstract}

In the framework of Einstein-Yang-Mills theories, we study the gauge Lorentz group and establish a particular correspondence between this case and a certain class of theories with torsion within Riemann-Cartan space-times. This relation is specially useful in order to simplify the problem of finding exact solutions to the Einstein-Yang-Mills equations. The applicability of the method is divided into two approaches: one associated with the Lorentz group $SO(1,n-1)$ of the space-time rotations and another one with its subgroup $SO(n-2)$. Solutions for both cases are presented by the explicit use of this correspondence and, interestingly, for the last one by imposing on our ansatz the same kind of rotation and reflection symmetry properties as for a nonvanishing space-time torsion. Although these solutions were found in previous literature by a different approach, our method provides an alternative way to obtain them and it may be used in future research to find other exact solutions within this theory.
\end{abstract}


\maketitle

\section{Introduction}

Research of the Einstein-Yang-Mills (EYM) model has shown it to be a field of successful results. In the same way that we can find solutions in General Relativity (GR) with Abelian gauge bosons \cite{Stephani}, we can also find more general solutions in presence of non-Abelian vector fields with a large number of interesting properties, despite the nonhair conjecture \cite{Volkov-Galt'sov}. The first non-Abelian solution in the presence of curved space-time was found numerically by Bartnik and McKinnon in the four-dimensional static spherically symmetric EYM-SU(2) theory \cite{Bartnik-McKinnon}. It is a particlelike system, unlike the Abelian case given by the U(1) gauge group of the Einstein-Maxwell theory, where such a distribution is prohibited, but the same EYM model does also contain a black hole configuration \cite{Bizon}.

Increasing the number $n$ of dimensions of the space-time, new exact solutions for the EYM-$SO(n-2)$ case were found by the Wu-Yang ansatz \cite{Habib-Halilsoy}. In our work, we arrive to the same result by making use of a spin connection-like ansatz with Yang-Mills (YM) charge and applying the standard class of symmetry conditions as those assigned to the fundamental geometrical quantities of a Riemann-Cartan (RC) manifold (i.e., curvature and torsion).

From a mathematical point of view, any gauge field over a pseudo-Riemannian manifold $\mathcal{M}$ (i.e., coupled to gravity) is associated with a Lie group $\mathcal{G}$ and is expressed by a connection 1-form $A$ in the principal bundle $\mathcal{P}\left(\mathcal{M},\mathcal{G}\right)$, which takes values on the Lie algebra. This gauge connection defines a covariant derivative on the tangent bundle of $\mathcal{G}$ and the subsequent 2-form gauge curvature $F$, which constitutes the physical field playing the role of carrier of an interaction (i.e., the YM field if $\mathcal{G}$ is a non-Abelian Lie group) \cite{Yang-Mills}:

\begin{equation}D_{\mu}=\nabla_{\mu}+\frac{i}{\sigma}\left[A_{\mu},\cdot \, \right]\,,\end{equation}

\begin{equation}F_{\mu \nu}=\partial_{\mu}A_{\nu}-\partial_{\nu}A_{\mu}+\frac{i}{\sigma}\left[A_{\mu},A_{\nu}\right]\,,\end{equation}
where $\sigma$ is related to the coupling constant.

Then, the following commutating relation is satisfied:

\begin{equation}\left[D_{\mu}, D_{\nu}\right]\,v^{\lambda}=R^{\lambda}\,_{\rho \mu \nu}\,v^{\rho}+\frac{i}{\sigma}\left[F_{\mu \nu},v^{\lambda}\right]\,,\end{equation}
where $R^{\lambda}\,_{\rho \mu \nu}=\partial_{\mu}\Gamma^{\lambda}_{\rho \nu}-\partial_{\nu}\Gamma^{\lambda}_{\rho \mu}+\Gamma^{\lambda}_{\omega \mu}\Gamma^{\omega}_{\rho \nu}-\Gamma^{\lambda}_{\omega \nu}\Gamma^{\omega}_{\rho \mu}$ are the components of the Riemann tensor derived from the Levi-Civita connection and $v^{\lambda}$ is an arbitrary vector.

Their behavior under a gauge transformation $S \in \mathcal{G}$ allows us to construct minimal coupling actions. In terms of their components, it is given by the following rules:

\begin{equation}A_{\mu} \rightarrow A_{\mu}^{'}=S^{-1}A_{\mu}S-i\sigma S^{-1}\partial_{\mu}S\,,\end{equation}

\begin{equation}F_{\mu \nu} \rightarrow F_{\mu \nu}^{'}=S^{-1}F_{\mu \nu}S\,.\end{equation}

On the other hand, RC space-times incorporate the notion of torsion as the antisymmetric part of the affine connection on the manifold:

\begin{equation}T^{\lambda}\,_{\mu \nu}=2\tilde{\Gamma}^{\lambda}\,_{[\mu \nu]}\,.\end{equation}

Note that the notation with tilde refers to elements defined within the RC manifold and with the absence of a tilde to elements defined within the torsion-free pseudo-Riemannian manifold. Additionally, according to the correspondence used by our method, the same convention applies to quantities depending on torsionlike components (i.e., corrections in the gauge potentials that are referred to internal symmetry groups and have similar algebraic symmetries in analogy to the torsion tensor).

Although the affine connection does not transform like a tensor under a general change of coordinates, its antisymmetric part does (i.e., torsion is a third-rank tensor and it cannot be locally vanished if it has not associated an absolute zero value). Furthermore, whereas curvature is related to the rotation of a vector along an infinitesimal path over the space-time, torsion is related to the translation and has deep geometrical implications, such as breaking infinitesimal parallelograms on the manifold \cite{Sab-Gas}.

Thus, unlike the torsion-free case where the geometry is completely described by the metric (i.e., the affine connection corresponds to the Levi-Civita connection), the presence of torsion introduces independent characteristics and modifies the expression of the affine connection in the following form:

\begin{equation}\tilde{\Gamma}^{\lambda}\,_{\rho \mu}={\Gamma}^{\lambda}\,_{\rho \mu}+{K}^{\lambda}\,_{\rho \mu}\,,\end{equation}
where ${K}^{\lambda}\,_{\rho \mu}=\frac{1}{2}(T^{\lambda}\,_{\rho \mu}-T_{\mu}\,^{\lambda}\,_{\rho}+T_{\rho \mu}\,^{\lambda})$ is the so-called contortion tensor and fulfills ${K}^{\lambda}\,_{\rho \mu}=-\,{K}_{\rho}\,^{\lambda}\,_{\mu}$, in order to preserve the metricity condition $\tilde{\nabla}_{\lambda}\,g_{\mu \nu}=0$ (i.e., the total covariant derivative of the metric tensor vanishes identically).

One of the most fundamental aspects of introducing these new geometrical characteristics within a physical theory of space-time and matter beyond GR is its main role as a dynamical field if higher order curvature and torsion terms are included in the Lagrangian. Whereas the so-called Einstein-Cartan theory only incorporates first-order corrections in the Lagrangian and therefore no propagating torsion is allowed, second-order corrections describe a Lagrangian with dynamical torsion depending on ten parameters \cite{Hehl, Obukhov}.

In the present work, we use these notions about the EYM theory and the quadratic gravitation theory with propagating torsion to bridge the gap between both in a very special case. Indeed, under a simple class of additional restrictions, we shall see that our assumptions allow us to obtain different classes of exact solutions to the EYM equations and to study other possible configurations in such a case. In this sense, the primary starting point of our analysis is based on the study of noncompact Lie groups. Although these constructions are related to nonunitary theories, one interesting aspect of this type of group is the possibility of establishing a correspondence between the theory under study and a set of modified theories of gravity with propagating space-time torsion, as is developed in this work. Indeed, the standard theory of gravity and the larger part of its extensions belong to this group. Following our discussion, we establish original dynamical constraints in order to simplify and to classify all the possible solutions derived by the approach described in the manuscript. 

This paper is organized as follows. Section II presents the general EYM-Lorentz field equations, as well as these equations under the spin connection-like ansatz and its association with a particular quadratic gravitational theory of second order in the curvature term with dynamical torsion. The general expressions for the metric and the torsion tensor under rotations and reflections in the static spherically symmetric space-time are shown in Section III. We apply these particular conditions and find the respective solutions for the torsionlike and torsionless cases in Section IV. Finally, the conclusions obtained from our analysis are presented in Section V.

\section{EYM-LORENTZ ANSATZ AND CONDITIONS}

We will use Planck units throughout this work ($G=c=\hbar=1)$ and consider for our study the following Lagrangian:

\begin{equation}S=-\frac{1}{16\pi}\int \left({R} - tr\,F_{\mu \nu} F^{\mu \nu}\right) \sqrt{-g}\,d^{n}x\,,\end{equation}
where the minimal coupling is assumed. Note that depending on the character of the gauge formalism and its corresponding group of transformations assumed by the approach, this action can be framed either on a modified gravity model or on a system of interaction between gauge fields and regular gravity. Specifically, gauging external or internal degrees of freedom is related to a large class of gauge gravity models based on space-time symmetries and to YM theories, respectively. In the present case, we consider both analyses with the external $SO(1,n-1)$ group and the internal $SO(n-2)$, in order to obtain a class of general constraints that allows us to classify their possible solutions under the appropriate correspondence conditions.

Therefore, the general equations derived from this action by performing variations with respect to the metric tensor and the gauge connection of the groups under consideration are:

\begin{equation}\label{gaugeeq}(D_{\mu}\,F^{\mu\nu})^{a b}=0\,,\end{equation}

\begin{equation}\label{einsteineq}
G_{\mu\nu}=8\pi T_{\mu\nu}\,,
\end{equation}
where $G_{\mu \nu} = R_{\mu\nu}-\frac{R}{2}g_{\mu\nu}$ is the Einstein tensor and $T_{\mu\nu} = \frac{1}{4\pi}\,tr\,\left(\frac{1}{4}g_{\mu\nu}F_{\lambda\rho}F^{\lambda\rho}-F_{\mu\rho}F_{\nu}\,^{\rho}\right)$, whereas latin a, b and greek $\mu$, $\nu$
indices run from $0$ to $n-1$ and refer to anholonomic and coordinate basis, respectively. Furthermore, the divergencelessness of the Einstein tensor implies the following conservation law:

\begin{equation}\label{divergencelesseq}
\nabla_{\mu}T^{\mu \nu}=0\,.
\end{equation}

These field equations typically constitute a complicated nonlinear system of equations and additional constraints are usually required in order to simplify the problem and to focus on particular cases. Then, by taking into account these lines, we assign the following spin connection-like ansatz to the gauge connection:

\begin{equation}\label{ansatz}
A^{a b}\,_{\mu}=Q\,\left(e^{a}\,_{\lambda}\,e^{b \rho}\,\tilde{\Gamma}^{\lambda}\,_{\rho \mu}+e^{a}\,_{\lambda}\,\partial_{\mu}\,e^{b \lambda}\right)\,.
\end{equation}

This expression usually represents a spin connection on a RC space-time (i.e., a curved space-time with torsion), so it can be regarded as the gauge field generated by local Lorentz transformations in such a case. Alternatively, under the EYM framework associated with internal gauge groups, it is always possible to select any particular ansatz in order to describe the respective YM field, so in this formalism we will start from the same mathematical expression and find embedded non-Abelian $SO(n-2)$ solutions.

The gauge connection can be written as $A_\mu=A^{a b}\,_{\mu}\,J_{a b}$, where $J_{ab}=i[\gamma_a,\gamma_b]/8$ are the generators of the Lorentz gauge group, which satisfy the following commutative relations:

\begin{equation}
[J_{ab},J_{cd}]=\frac{i}{2}\,(\eta_{ad}\,J_{bc}+\eta_{cb}\,J_{ad}-\eta_{db}\,J_{ac}-\eta_{ac}\,J_{bd})\,.
\end{equation}

By using the antisymmetric property of the gauge connection with respect to
the Lorentz indeces, $A^{a b}\,_{\mu}=-\,A^{b a}\,_{\mu}$, we can write the field strength tensor
as
\begin{equation}
F^{a b}\,_{\mu\nu}=\partial_{\mu} A^{a b}\,_{\nu}-\partial_{\nu} A^{a b}\,_{\mu}+\frac{1}{\sigma}\left(A^{a}\,_{c \mu}\,A^{c b}\,_{\nu}-A^{a}\,_{c \nu}\,A^{c b}\,_{\mu}\right)\,.
\end{equation}

Finally, by taking into account the orthogonal property of the tetrad field $e_{a}\,^{\lambda}\,e^{a}\,_{\rho}=\delta_{\rho}^{\lambda}$ and setting $\sigma = Q$,
the field strength tensor takes the form \cite{Utiyama,Yepez}:
\begin{equation}\label{strength}
F^{a b}\,_{\mu\nu}=Q\,e^{a}\,_{\lambda}\,e^{b}\,_{\rho}\,\tilde{R}^{\lambda \rho}\,_{\mu \nu}\,,
\end{equation}
where $\tilde{R}^{\lambda}\,{_{\rho \mu \nu}}$ coincides with the general expression of the components of the Riemann tensor over a RC space-time.

Rewriting the above action under the spin connection-like ansatz, it turns out that it coincides with the following quadratic gravity action in presence of torsion:

\begin{equation}S=-\frac{1}{16\pi}\int \left({R} - 2^{\tilde{n}/2-3}Q^{2}\,\tilde{R}_{\lambda \rho \mu \nu} \tilde{R}^{\lambda \rho \mu \nu}\right) \sqrt{-g}\,d^{n}x\,,\end{equation}
with $\tilde{n}=n$ and $\tilde{n}=n-1$ for even and odd $n$.

Therefore, Eqs. (\ref{gaugeeq}) and (\ref{einsteineq}) for such a case can be expressed in terms of geometrical quantities, respectively, as follows:

\begin{equation}\label{gaugeeq2}
\partial_{\rho}\tilde{R}_{\mu}\,^{\lambda \nu \rho}+\Gamma^{\rho}\,_{\omega \rho}\tilde{R}_{\mu}\,^{\lambda \nu \omega}+\tilde{\Gamma}^{\lambda}\,_{\omega \rho}\tilde{R}_{\mu}\,^{\omega \nu \rho}-\tilde{\Gamma}^{\omega}\,_{\mu \rho}\tilde{R}_{\omega}\,^{\lambda \nu \rho}=0\,,
\end{equation}

\newcommand*\rfrac[2]{{}^{#1}\!/_{#2}}

\begin{equation}\label{eq3}
R_{\mu\nu}-\frac{R}{2}g_{\mu\nu}=2^{\rfrac{\tilde{n}}{2}}Q^{2}\left(g_{\mu \nu}\tilde{R}_{\lambda \rho \omega \tau}\tilde{R}^{\lambda \rho \omega \tau}-4\tilde{R}^{\lambda \rho}\,_{\mu \omega}\tilde{R}_{\lambda \rho \nu}\,^{\omega}\right)\,.
\end{equation}

Thus, if a certain class of space-time symmetries are imposed, then not only the condition $\mathcal{L}_{\xi}g_{\mu \nu}=0$ must be satisfied, but also $\mathcal{L}_{\xi}T^{\lambda}\,_{\mu \nu}=0$ (i.e., the Lie derivative in the direction of the Killing field $\xi$ on $T^{\lambda}\,_{\mu \nu}$ vanishes), in order to preserve the reasonable curvature and torsion symmetries.

\section{SPHERICAL AND REFLECTION SYMMETRIES}

The metric of a $n$-dimensional static spherically symmetric space-time can be written as

\begin{equation}ds^2=A(r)\,dt^2-\frac{dr^2}{B(r)}-r^2d\Omega^{2}_{n-2}\,,\end{equation}
where $d\Omega_{n-2} = d\theta_{1}^{2}+\sum_{i=2}^{n-2}\prod_{j=1}^{i-1}\sin^{2}\theta_{j}d\theta_{i}^{2}$, with $0 \leq \theta_{n-2} \leq 2\pi$ and $0 \leq \theta_{k} \leq \pi$, $1 \leq k \leq n-3$. We assume $n\geq 4$.

Then, it can be shown that the only nonvanishing components of $T^{\lambda}\,_{\mu \nu}$ are \cite{Rauch-Nieh,Sur-Bhatia}:

\begin{eqnarray}
T^{t}\,_{t r}&=&a(r) \;\;\, ;\nonumber\\
T^{r}\,_{t r}&=&b(r) \;\;\; ;\nonumber\\
T^{\theta_{k}}\,_{t \theta_{l}}&=&\delta^{\theta_{k}}\,_{\theta_{l}}\,c(r) \;\;\; ;\nonumber\\
T^{\theta_{k}}\,_{r \theta_{l}}&=&\delta^{\theta_{k}}\,_{\theta_{l}}\,g(r) \;\;\; ;\nonumber\\
T^{\theta_{k}}\,_{t \theta_{l}}&=&e^{a \theta_{k}}\,e^{b}\,_{\theta_{l}}\,\epsilon_{a b}\, d (r)\,, \;\;\;\;\;\;\;\;\;\;\;\;\;\;\;\;\; \text{if}\;\;\; n=4\;\;\; ;  \nonumber\\
T^{\theta_{k}}\,_{r \theta_{l}}&=&e^{a \theta_{k}}\,e^{b}\,_{\theta_{l}}\,\epsilon_{a b}\, h (r)\,, \;\;\;\;\;\;\;\;\;\;\;\;\;\;\;\;\; \text{if}\;\;\; n=4\;\;\; ;  \nonumber\\
T^{t}\,_{\theta_{k} \theta_{l}}&=&\epsilon_{k l} k (r)\,\sin\theta_{1}\,, \,\;\;\;\;\;\;\;\;\;\;\;\;\;\;\;\;\;\;\;\;\; \text{if}\;\;\; n=4\;\;\; ; \nonumber\\
T^{r}\,_{\theta_{k} \theta_{l}}&=&\epsilon_{k l} l (r)\,\sin\theta_{1}\,, \,\;\;\;\;\;\;\;\;\;\;\;\;\;\;\;\;\;\;\;\;\;\, \text{if}\;\;\; n=4\;\;\; ; \nonumber\\
T^{\theta_{k}}\,_{\theta_{l} \theta_{m}}&=&\,e^{a \theta_{k}}\,e^{b}\,_{\theta_{l}}\,e^{c}\,_{\theta_{m}}\,\epsilon_{a b c}\, f (r)\,,\;\;\;\;\;\;\;\, \text{if}\;\;\; n=5 \;\;\;,
\end{eqnarray}
where $a,b,c,d,g,h,k$ and $l$ are arbitrary functions depending only on r; $k,l=1,2$, and $\epsilon_{a b}$, $\epsilon_{a b c}$ are the totally antisymmetric Levi-Civita symbol of second and third order, respectively.

Therefore, in addition to the two functions associated with the metric, for $n=4$ dimensions, there are still a total number of eight unknown independent functions to solve the field equations. Furthermore, imposing reflection symmetry (i.e., $O(3)$ spherical symmetry) requires that $d(r)$, $h(r)$, $k(r)$ and $l(r)$ vanish, so that the number reduces to four.

\section{SOLUTIONS}

In order to categorize all the possible solutions, we can rewrite Eq. (\ref{gaugeeq2}) in the following form:

\begin{equation}
\nabla_{\rho}R_{\mu}\,^{\lambda \nu \rho}+\nabla_{\rho}T_{\mu}\,^{\lambda \nu \rho}+K^{\lambda}\,_{\omega \rho}\tilde{R}_{\mu}\,^{\omega \nu \rho}-K^{\omega}\,_{\mu \rho}\tilde{R}_{\omega}\,^{\lambda \nu \rho}=0\,,
\end{equation}
where $T^{\lambda}\,_{\rho \mu \nu} = \nabla_{\mu}K^{\lambda}\,_{\rho \nu}-\nabla_{\nu}K^{\lambda}\,_{\rho \mu}+K^{\lambda}\,_{\sigma \mu}K^{\sigma}\,_{\rho \nu}-K^{\lambda}\,_{\sigma \nu}K^{\sigma}\,_{\rho \mu}$ coincides with the torsion contribution to the curvature tensor of the RC space-time, so that we can distinguish between the torsion-free and the torsion parts if it is required.

On the other hand, according to the second Bianchi identity for a pseudo-Riemannian manifold, the components of the Riemann tensor in such a manifold satisfy:

\begin{equation}
{\nabla}_{[\lambda |}R^{\omega}\,_{\rho | \mu \nu]}=0\,.
\end{equation}

By contracting this expression with the metric tensor and considering the above form of the mentioned field equation, it is straightforward to obtain the following condition for our model:

\begin{equation}\label{ec1}2\nabla_{[\mu}R_{\nu]\rho}=\nabla_{\lambda}T_{\mu \nu \rho}\,^{\lambda}+2K^{\omega}\,_{[\mu | \lambda}\tilde{R}_{\nu ] \omega \rho}\,^{\lambda}\,.\end{equation}

In addition, the conservation law (\ref{divergencelesseq}) turns out to be equivalent to the following expression:

\begin{equation}\label{ec2}\frac{1}{2}\nabla_{\nu}R+\nabla_{\lambda}T_{\mu \nu}\,^{\mu \lambda}+K^{\omega}\,_{\mu \lambda}\tilde{R}_{\nu \omega}\,^{\mu \lambda}-K^{\omega}\,_{\nu \lambda}\tilde{R}_{\omega}\,^{\lambda}=0\,.\end{equation}

These expressions are shown as generic conditions of this model and they will allow us to classify all the possible configurations in the most important cases.

Before distinguishing between torsionless and nonvanishing torsionlike cases, let us summarize the respective assumptions that allow us to establish and to obtain the distinct classes of solutions according to our discussion. The starting point is the mapping defined in Eq. (\ref{ansatz}), which coincides with the well-known spin connection of a given space-time. This quantity has typically been used in order to describe appropriately the dynamics of the fermion fields on a general space-time. It has also been used in the most important gauge theories of gravity, such as the well-known Lorentz gauge gravity or the Poincar\'e gauge gravity, since it gives rise to a Lorentz gauge curvature which is proportional to the Riemann tensor, as is shown in Eq. (\ref{strength}).

Continuing with our analysis, when the nonvanishing torsionlike $O(n-1)$ symmetric and the purely magnetic cases are considered in a $n$-dimensional static and spherically symmetric space-time, the system of equations given by Eq. (\ref{gaugeeq2}) and Eq. (\ref{eq3}) together with the constraints (\ref{ec1}) and (\ref{ec2}) will allow us to find the mentioned embedded $SO(n-2)$ solutions. It is straightforward to check the dimension of this gauge group by computing the independent connection components of the solutions, giving rise to a dimension of $(n-2)(n-3)/2$, as expected.

\subsection{Torsionless case}

For the torsionless $SO(1,n-1)$ case, the following constraint is satisfied:

\begin{equation}\nabla_{[\lambda}R_{\rho]\nu}=0\,,\end{equation}
with:

\begin{equation}\left[\nabla_{[\mu},\nabla_{\nu |}\right]R_{\lambda |\rho]}=-R_{[\mu \nu| \lambda}\,^{\omega}R_{|\rho] \omega}\,.\end{equation}

Thus, the existence of the integrability condition $R_{[\mu \nu|\lambda}\,^{\omega}R_{|\rho] \omega}=0$ allows us to solve this equation and obtain the following solutions \cite{LsTr}:

\begin{equation}
\label{ec}R_{\mu \nu}=b\,g_{\mu \nu}\,,
\end{equation}
where $b$ is a constant.

Therefore, the only possible geometries for this torsionless case correspond to Einstein manifolds. Note that the tracelessness of the torsion-free Einstein tensor in four dimensions implies that $b=0$, so these solutions satisfy $R_{\mu \nu}=0$ (i.e., the space-time is Ricci-flat). On the other hand, by increasing the number of dimensions, the corrections to the gravitational field act as a cosmological constant in the Einstein equations \cite{Cemb-Gig}.

\subsection{Nonvanishing torsionlike case}

The condition (\ref{ec1}) equal to zero enables the existence of Einstein manifold solutions even for the case of an external symmetry group $SO(1,n-1)$ in the presence of a nonvanishing space-time torsion. However, other geometries are allowed according to the generic conditions (\ref{ec1}) and (\ref{ec2}).

Particularly, for a $n$-dimensional static spherically symmetric space-time, if we simplify the problem using the previous considerations and restrict to the internal gauge group $SO(n-2)$, it is possible to find the following purely magnetic black hole solutions to the resulting EYM equations with $O(n-1)$ symmetric torsionlike tensor (rotation and reflection symmetric):

\begin{eqnarray}\label{ec3}
T^{t}\,_{t r}&=&\frac{A'(r)}{2A(r)} \,,\;\;
T^{\theta_{k}}\,_{r \theta_{k}}=-\,\frac{1}{r} \,,\;\;
T^{r}\,_{t r}=T^{\theta_{k}}\,_{t \theta_{k}}=T^{\theta_{k}}\,_{\theta_{l} \theta_{m}}=0 \;;
\end{eqnarray}
with

\begin{equation}\label{ec4}
A(r)=B(r)= \left\{
\begin{array}{l}
1-\frac{2m}{r^{2}}-\frac{2Q^{2}ln(r)}{r^2}\,, \;\;\;\;\;\;\;\;\;\;\;\;\;\, \text{if} \;\;\; n=5 \\
  \\
1-\frac{2m}{r^{n-3}}-2^{\tilde{n}/2-2}\frac{(n-3)Q^2}{(n-5)r^2}\,, \;\;\, \text{if} \;\;\; n \neq 5\,. \\
\end{array}
\right.
\end{equation}

Although these geometries are asymptotically flat, for $n=5$ and $n \geq 6$ dimensions their Arnowitt-Deser-Misner (ADM) mass \cite{ADM} diverges as $ln(r)$ and $r^{n-5}$, respectively. Nevertheless, solutions with finite ADM mass are found by including higher order terms of the YM hierarchy in the Lagrangian \cite{Radu-Tchrakian, RRadu-Tchrakian}.

The nonvanishing components of the field strength tensor are:

\begin{equation}
F^{a b}\,_{\theta_{i} \theta_{j}}=Q\,e_{\theta_{i}}\,^{a}\,e_{\theta_{j}}\,^{b}\,\tilde{R}^{\theta_{i} \theta_{j}}\,_{\theta_{i} \theta_{j}}\,,
\end{equation}
with $\tilde{R}^{\theta_{i} \theta_{j}}\,_{\theta_{i} \theta_{j}}=-\frac{1}{r^2}$.

For $n = 4$ dimensions, the system reduces to the EYM-$SO(2)$ case, which is indeed equivalent to the magnetic Einstein-Maxwell solution, because of the isomorphism between $SO(2)$ and the $U(1)$ group. On the other hand, for $n \geq 5$ dimensions the existence of these EYM-$SO(n-2)$ solutions describes the coupling of a nontrivial YM magnetic field to gravity.

It has also been shown by different ways that these solutions have a number of interesting properties and they are compatible with the existence of a cosmological constant and Maxwell fields, as well as with other modified theories of gravity, such as Gauss-Bonnet gravity \cite{Habib-Halilsoy, HHabib-Halilsoy}.

This work completes our previous study on EYM theory presented in \cite{Cemb-Gig}. More general solutions may be found using this method, especially for $n=4$ dimensions since the $\mathcal{L}_{\xi}T^{\lambda}\,_{\mu \nu}=0$ condition allows a richer structure than for any other number of dimensions.

\section{DISCUSSION AND SUMMARY}

In this article, we have presented a new method to find exact solutions to the EYM-Lorentz theory, based on the correspondence between the EYM system and a certain class of quadratic gravity theories in the presence of torsion, under the restriction introduced by the spin connection-like ansatz. The available configurations can be categorized into the torsionless and the nonvanishing torsionlike cases, according to general conditions. For the torsionless branch, it is shown that the only possible geometries correspond to Einstein manifolds associated with the external group $SO(1,n-1)$, whereas for the nonvanishing torsionlike branch the method allows us to distinguish the mentioned external group of symmetries from the internal $SO(n-2)$ and other families of embedded solutions emerge. These solutions describe a sort of purely magnetic black hole with YM charge and they were found earlier by different approaches \cite{Habib-Halilsoy, HHabib-Halilsoy}.

Note that these results are derived from similar mathematical expressions, but they refer to completely different approaches. Namely, from a gauge-theoretical approach, it is a well-known fact that the presence of a space-time torsion requires gauging the external degrees of freedom consisting of rotations and translations, in a way that both curvature and torsion are inexorably related to the rotation and the translation noncompact groups, respectively \cite{Hehl, Obukhov}. Furthermore, as previously stressed, the displacement of a vector along an infinitesimal path in a RC manifold involves a breaking of the consequent parallelograms defined on such a manifold, in a way that its translational closure failure proportionally depends on the torsion tensor \cite{Sab-Gas}. Therefore, the embedding of the $SO(n-2)$ group corresponds to a distinct configuration where the resulting gauge connections are accordingly related to an internal symmetry group and the additional torsionlike degrees of freedom contained in the latter do not represent a space-time torsion but a third-rank tensor with similar algebraic symmetries that provides a purely magnetic black hole solution to the variational equations. Indeed, it is straightforward to check from the nonvanishing torsionlike components of this solution that the corresponding  $SO(n-2)$ gauge connection and its associated field strength tensor can be written as $A_\mu=A^{\tilde{a} \tilde{b}}\,_{\mu}\,J_{\tilde{a} \tilde{b}}$ and $F_{\mu \nu}=F^{\tilde{a} \tilde{b}}\,_{\mu \nu}\,J_{\tilde{a} \tilde{b}}$, respectively, with $\tilde{a}, \tilde{b}=2, ..., n-1$. Thus, it is clear that these quantities are connected to the mentioned gauge group instead of an external symmetry group related to the space-time rotations or translations.

On the other hand, further implications arise when considering the coupling with matter fields. For instance, if we study the dynamics of a Dirac fermion within the solution given by Eqs. (\ref{ec3}) and (\ref{ec4}), the behavior is completely different than that which occurred in the presence of a space-time torsion, where the fermion would irremediably suffer the associated spin connection. However, in the first case, the fermion would interact with the $SO(n-2)$ gauge interaction depending on its particular multiplet representation (for $n>4$) or charge (for $n=4$). In the simplest case, it could even be a singlet ($n>4$) or neutral (zero charge), so it would not interact with the new gauge force.

This fact contrasts with some publications that do not bear in mind these fundamental relations and wrongly try to identify the space-time torsion with YM or electromagnetic fields (see Fallacy 9 on page 267 of reference \cite{Blag-Hehl}). Thus, our $SO(n-2)$ solution is not covered by this sort of fallacy, in the same way as the Mazharimousavi-Halilsoy solution since both solutions coincide and represent the same type of configuration.

Finally, it is worthwhile to stress that distinct classes of EYM-Lorentz systems that are physically meaningful may be found using our ansatz, especially in $n=4$ dimensions because the $\mathcal{L}_{\xi}T^{\lambda}\,_{\mu \nu}=0$ condition could allow for more complex solutions. Additionally, for the development of this aim, the general condition $\tilde{\nabla}_{\lambda}\,g_{\mu \nu}=0$ still holds, but it could be also possible to deal with the same analysis relaxing this restriction, in order to find different EYM systems related to this geometrical property. Within this framework, an interesting and simple case might arise from the Weyl-Cartan geometry, where the nonmetricity condition is expressed as $\tilde{\nabla}_{\lambda}\,g_{\mu \nu}=w_{\lambda}\,g_{\mu \nu}$, so that the number of irreducible decomposition pieces of nonmetricity reduces to the Weyl 1-form $w$ \cite{HHehl}.
Further research following these lines of study will be addressed in the future.

\bigskip
\bigskip
\noindent
{\bf ACKNOWLEDGMENTS}

\bigskip

We would like to thank Luis J. Garay and Antonio L. Maroto for helpful discussions.
This work has been supported in part by the MINECO (Spain) project Nos. FIS2014-52837-P, FPA2014-53375-C2-1-P, and Consolider-Ingenio MULTIDARK Grant. No. CSD2009-00064.
J.A.R.C. acknowledges financial support from a Jose Castillejo grant (2015). 
J.G.V. acknowledges support from a MULTIDARK summer student fellowship (2015).

\end{document}